\documentclass[prd,a4paper,showpacs,preprintnumbers,nofootinbib]{revtex4} 

\newcommand{\bm}[1]{\mbox{\boldmath $#1$}}

\newcommand{\open}{{<\kern -0.3 em{\scriptscriptstyle )}}}

\newcommand{\nslash}{\kern 0.2 em n\kern -0.45em /}
\newcommand{\Pslash}{\kern 0.2 em P\kern -0.56em \raisebox{0.3ex}{/}}
\newcommand{\pslash}{\kern 0.2 em p\kern -0.4em /}
\newcommand{\kslash}{\kern 0.2 em k\kern -0.45em /}
\newcommand{\Sslash}{\kern 0.2 em S\kern -0.56em \raisebox{0.3ex}{/}}
\newcommand{\slsh}[1]{\mbox{$\not\! #1$}}

\newcommand{\eq}{\begin{equation}}
\newcommand{\ee}{\end{equation}}
\newcommand{\beq}{\begin{equation}}
\newcommand{\eeq}{\end{equation}}
\newcommand{\ba}{\begin{eqnarray}}
\newcommand{\ea}{\end{eqnarray}}
\newcommand{\eqa}{\begin{eqnarray}}
\newcommand{\eea}{\end{eqnarray}}

\newcommand{\psibar}{\overline{\psi}}

\newcommand{\sumint}{\kern 0.2 em {\textstyle\sum} \kern -1.1 em \int}

\newcommand{\la}{\langle}
\newcommand{\ra}{\rangle}
\newcommand{\amp}[1]{\la #1 \ra}

\begin{document} 

\title{On a possible node in the Sivers and Qiu-Sterman functions}

\author{Dani\"el Boer}
\email{D.Boer@rug.nl}
\affiliation{Theory Group, KVI, University of Groningen,
Zernikelaan 25, 9747 AA Groningen, The Netherlands}

\date{\today}

\begin{abstract}
The possibility of a node in the $x$ dependence of the Sivers and Qiu-Sterman functions is 
discussed in light of its importance for the experimental check of the overall sign change of 
the Sivers effect between semi-inclusive DIS and the Drell-Yan process.   
An $x$-dependent version of the Ehrnsperger-Sch\"afer-Greiner-Mankiewicz relation between 
the Qiu-Sterman function and a twist-3 part of $g_2$ is presented, which naturally suggests a 
node in the Qiu-Sterman function. This relation could be checked experimentally 
as well and could provide qualitative information on the gluonic field strength inside 
the proton. Satisfying the Burkardt sum rule by means of a node is briefly discussed and the importance
of modelling the Sivers function including its full Wilson line is pointed out.
\end{abstract}


\maketitle


The Sivers \cite{Sivers} and Qiu-Sterman \cite{QS} effects have
been proposed as possible explanations of single transverse spin
asymmetries $A_N$ observed in the process $p^\uparrow \, p \to \pi \, X$ 
\cite{Adams:1991cs}. In recent years it has become clear that these two effects are 
intimately related \cite{Boer:2003cm,Ji:2006ub}. In this paper the $x$ dependence 
of these effects is discussed, in particular the possibility of a node. 

The Sivers effect is described by a transverse momentum 
dependent parton distribution function (TMD) and generates azimuthal
spin asymmetries for instance in the Drell-Yan (DY) process and in semi-inclusive
DIS (SIDIS). The Sivers effect asymmetry in SIDIS has been clearly observed 
in the HERMES \cite{Airapetian:2009ti} and COMPASS \cite{Alekseev:2010rw} experiments. 
The Sivers function, here denoted by $f_{1T}^\perp(x,k_T^2)$, describes the 
difference between the probability to find a quark with lightcone momentum fraction 
$x$ and transverse momentum $k_T$  inside a hadron polarized transversely to its 
momentum direction and the one where the polarization points in the opposite direction.
As the Sivers function describes a difference of probabilities it is not necessarily positive definite.
In fact, the major interest in extracting the Sivers function from the DY process is that it is 
expected to have the opposite sign compared to the one extracted from SIDIS \cite{Collins:2002kn}. 
The gauge invariant definition of the Sivers function is in terms of a nonlocal operator involving a 
Wilson line:
\beq
f_{1T}^{\perp [{\cal C}]}(x,k_T^2)\,  \epsilon_T^{\alpha \beta} S_{T\alpha} k_{T\beta} = 
\frac{M}{2}\,  {\rm F.T.} \left. \amp{P, S_T|\, \psibar(0) \, 
{\cal L}_{\cal C}(0,\xi) \, \slsh{n_-} \, \psi(\xi) \,|P, S_T}\right|_{\xi^+=0},
\eeq
where ${\cal L}_{\cal C}$ denotes the Wilson line along contour ${\cal C}$; $S_T$ denotes the transverse spin vector; 
and F.T. denotes taking the Fourier transform, where 
$\xi^-$ and $\xi_T$ are the Fourier conjugate variables of $xP^+$ and $k_T$, respectively. The Sivers function is not 
uniquely defined, as it depends on the contour of the Wilson line, which in turn depends on the 
process considered. The Sivers function appearing in SIDIS contains a future 
pointing Wilson line, whereas in DY it is the same except past pointing, leading to the following overall sign relation \cite{Collins:2002kn}:
\beq
f_{1T}^{\perp [{\rm SIDIS}]}(x,k_T^2) = - f_{1T}^{\perp [{\rm DY}]}(x,k_T^2). 
\eeq
This is a {\it prediction\/} of the TMD formalism that remains to be tested. A related sign test 
in $W$ and $Z$ production at RHIC has been put forward in Refs.\ \cite{Kang:2009bp,Kang:2009sm}. 
In more complicated processes that allow TMD factorization, Sivers functions with other Wilson lines can appear, 
which are not simply related by an overall sign to the Sivers function of SIDIS to which we will refer as ``the" Sivers function 
from now on. 

It is important to emphasize that the above sign relation is about the overall sign and that the Sivers function itself 
need not be of fixed sign as a function of $x$. It can have one or more nodes in the $x$ dependence, even though 
its present extraction from SIDIS data in a restricted $x$ range does not display a node \cite{Anselmino:2008sga}. 
Nevertheless, the possibility of a node should be kept in mind when comparing the extraction from SIDIS with the 
future one from DY. A node position is generally expected to be $Q^2$ dependent, therefore, unless one compares the 
functions at the same $x$ and $Q^2$ values, the sign change test need not be conclusive. 
Moreover, such a node need not be at the same position for the different flavors, possibly complicating the 
comparison further. Unless one makes sure that nodes do not play a role in the comparison, one may 
wrongly jump to the conclusion that the TMD formalism is flawed in some way if the overall sign change between SIDIS
and DY is not confirmed in experiment. Motivated by the importance to know whether the Sivers function has 
a node, we investigate if there are any other indications in favor or against such a node. 

Since we are interested in the $x$ dependence here, 
we will not address the transverse momentum dependence (for a discussion of possible nodes in the $k_T$
dependence cf.\ \cite{Kang:2011hk}) and restrict to the first transverse
moment of the Sivers function, i.e.\ the Sivers function weighted with
the transverse momentum squared:
\beq
f_{1T}^{\perp (1)}(x) \equiv \int d^2 \bm{k}_T^{} \, 
\frac{\bm{k}_T^2}{2 M^2} \, f_{1T}^\perp(x,\bm{k}_T^2) .
\eeq 
This quantity is of interest because of its direct relation to the twist-3 Qiu-Sterman (QS)
function $T(x,S_T)$ \cite{QS}:
\beq
{T(x,S_T)} = i \frac{M}{P^+} \int \frac{d \lambda}{2\pi} e^{i\lambda x}\amp{P,S|\,  \psibar (0) {\Gamma_\alpha} 
{\int d\eta \;  
F^{+\alpha} (\eta n_-)}\; \psi(\lambda n_-)\, | P,S},
\eeq
where ${\Gamma_\alpha} \equiv  \epsilon_{T \beta \alpha} S_T^{\beta}
\slsh{n_-}/(2i M P^+) $ (like in \cite{QS} we take $\vec{S}_T^2=1$) 
and $\epsilon_T^{\mu\nu}=\epsilon^{\alpha \beta\mu\nu} n_{+\alpha}
n_{-\beta}$. 
The relation between the first moment of the Sivers function (of SIDIS) and the QS function is 
a direct proportionality \cite{Boer:2003cm}:
\beq
f_{1T}^{\perp (1)}(x) = - \frac{g}{2M} T(x,S_T).
\eeq 
The $x$-dependence of the two functions is therefore the same, apart from an overall 
proportionality constant. From now on we will absorb the coupling constant $g$ into the definition 
of $T$, but it is displayed here explicitly since it determines the relative sign in front of it (for 
a discussion of this issue cf.\ \cite{Kang:2011hk}).

The above definition of the QS function is given in the
$A^+=0$ lightcone gauge for simplicity and contains the operator 
$\int d\eta \;  F^{+\alpha}(\eta n_-)$, which renders it intrinsically nonlocal along the 
lightcone, even in the $A^+=0$ gauge. Below we are going to discuss two 
different assumptions about this lightcone integral over the gluonic field strength, one by 
Qiu and Sterman \cite{QS} and one by Ehrnsperger, Sch\"afer, Greiner and 
Mankiewicz (ESGM) \cite{ESGM}. 

Qiu and Sterman considered the following parametrization \cite{QS}:
\beq
T^q(x,S_T) = \kappa_q\, \lambda\, f_1^q(x),
\label{QSparam}
\eeq
where $f_1(x)$ is the ordinary unpolarized parton distribution function
and $q$ denotes the quark flavor. This parametrization follows when the QS matrix element 
is viewed as yielding the average value of $\int d\eta \; F^{+\alpha}(\eta n_-)$ inside the unpolarized 
proton and hence is expected to be simply a number times the unpolarized distribution
function:
\beq
{f_1(x)}  =  \frac{1}{2P^+} \int \frac{d \lambda}{2\pi} e^{i\lambda x}\amp{P|\,  \psibar (0) \slsh{n_-} \psi(\lambda n_-)\, | P}.
\eeq
The function $f_1(x)$ is a probability distribution and hence of definite sign. 
Using this parametrization, $T(x,S_T)$ can have different signs for different flavors, but cannot 
exhibit any node. In order to roughly describe the single transverse spin asymmetries (SSA) experimentally
measured in the process $p^\uparrow \, p \to \pi \, X$ at $\sqrt{s} \approx 20$ GeV \cite{Adams:1991cs},  
the following choices were made for the parameters: $\kappa_u=+1=-\kappa_d$, $\kappa_s=0$, yielding 
$\lambda \sim 100$~MeV \cite{QS}. This parametrization was subsequently used to predict
SSA in $\pi$ production at $\sqrt{s} = 200$ GeV \cite{QS98} 
and in Drell-Yan \cite{Hammon:1996pw,Boer:2001tx}. No conclusive
evidence in favor of the above parameterization has been obtained yet
though. 

Another view on the lightcone integral over the gluonic field strength is taken by ESGM. In Ref.\ \cite{ESGM} 
it is assumed that the field strength is only significantly contributing inside the proton and that 
the following approximation holds: 
\beq
\int d \eta \; F^{+\alpha}(\eta n_-) \approx 
F^{+\alpha}(0) \times 2 c M R_0,
\label{ESGMapprox}
\eeq
where $R_0$ is the proton ``lightcone'' radius in the rest
frame of the proton (taking into account that it is being probed by a highly relativistic probe, which sees the 
proton as Lorentz contracted) and $V\equiv \int d \eta= 2 c M R_0$, where $c$ is argued to be a constant 
between 1/3 and 1 \cite{ESGM,Schafer:1992sx}. ESGM considered the field strength at $\eta =0$, because that 
corresponds to the position of the quark fields upon integration of $T(x,S_T)$ over $x$, such that a local operator is 
obtained.  
{\it A priori\/} it is not known whether this approximation is legitimate, except that it may simply be true numerically 
for some $A^\mu$ configurations, but of course there is no way to select such configurations. Nevertheless,
it leads to an interesting result, which is the ESGM relation between the
lowest (zeroth) Mellin moment of the QS function and the second
moment of the twist-3 part of the distribution function $g_2$, which according to \cite{ESGM} is:
\beq
\int_{-1}^1 {T(x,S_T)}\; dx = - 12 c M^2 R_0 \int_0^1 {x^2
\left. g_2(x)\right|_{\text{{\small twist-3}}}} \; dx.
\label{ESGM}
\eeq
It should be emphasized that this is {\em not\/} an exact relation. 
But if ESGM's approximation is fine, then one can relate
the magnitude of the above mentioned SSA to $g_2$; this would be very useful, since it is 
known that $\int x^2 \left. g_2(x)\right|_{\text{{\small twist-3}}} \,
dx$ is very small. 

The structure function $g_2$, which in the parton 
model is directly related to the distribution function $g_2^q$ via $g_2(x) = \frac{1}{2} \sum_{q,\bar{q}} e_q^2 g_2^q$,  
with $e_q^2$ the quark charge squared in units of the electron charge, 
has been measured by the E155 experiment at
SLAC. Also the twist-3 part of $g_2$ was extracted, yielding a value for its
second moment $d_2$ which is defined as 
\beq
d_2 = 3 \int_0^1 x^2
\left. g_2(x)\right|_{\text{{\small twist-3}}} dx 
\eeq
The E155 experiment obtained $d_2 = 0.0032 \pm 0.0017$ for the proton 
\cite{E155}, when taking into account all available SLAC data. 
Assuming there are no large cancellations among the quark
flavors, one concludes that also $d_2^u$ and $d_2^d$ are both very
small. This conclusion is supported by lattice QCD evaluations \cite{Gockeler:2000ja}. 

Combining the ESGM relation and the above parametrization (\ref{QSparam})
of the QS function would suggest a very small SSA in $p^\uparrow \, p \to \pi \, X$ 
contrary to observations. Given the very small size of $d_2$, one would conclude 
that either $\lambda$ is much smaller than expected from the SSA data or 
$R_0$ has to be unnaturally large ($\gg$ 1 fm). Failure of the ESGM relation is one possibility and 
may indicate some qualitative features of the lightcone integral of the gluonic 
field strength inside the proton, such as that it is not slowly varying inside the proton or is significantly 
contributing outside the proton too. But one could also question the 
validity of parametrization (\ref{QSparam}), because one 
way to allow for large SSA and simultaneously accomodate small $d_2$ through the ESGM 
relation is to consider the option that $T(x,S_T)$ changes sign as a function of $x$,
having large absolute value in certain $x$ regions, but having a small integral. This is the 
view we will explore here, leaving aside the idea that $T(x,S_T)$ is proportional to 
$f_1(x)$. 

The approximation (\ref{ESGMapprox}) can be extended in various ways 
to obtain an $x$-dependent version of the ESGM relation. For instance, 
assuming that the gluonic field strength is a slowly varying function (or constant even), 
within the proton, leads to the approximation 
\beq
\int d \eta \; F^{+\alpha}(\eta n_-) \approx 
F^{+\alpha}(\eta_0 n_-) \times 2 c M R_0,
\label{extendedapprox}
\eeq
for some (or any) $\eta_0$ within the proton lightcone radius $R_0$. To reduce to the ESGM relation
upon integration over $x$, one has to consider $\eta_0$ at the position of either one of the quark fields. This 
is automatically satisfied if the field strength is taken to be constant within the proton. It should be 
mentioned that if one views the above approximation as giving the average field strength times the integration 
region, that in that case the estimate of $c$ of Refs.~\cite{ESGM,Schafer:1992sx} may be far off if the gluonic 
field is in fact heavily fluctuating. This can be experimentally investigated. 

With the approximation (\ref{extendedapprox}) for $\eta_0=\lambda$, the following unintegrated relation 
can be obtained\footnote{To obtain this result the lightcone integral is implemented as centered
around zero, such that in the notation of \cite{Boer:1997bw} the correlator $\Phi_F^\alpha(x,y)$ is symmetric 
under the interchange of $x$ and $y$ and $\Phi_A^\alpha(x,y)$ antisymmetric and therefore $\tilde{G}_A(x,y)=0$.}:
\beq
{T^q(x,S_T)} = - 2 c M^2 R_0 \; x^2 \, \tilde{g}_T^q(x).
\label{unintegratedrelation}
\eeq
This is an unintegrated version of the ESGM relation, which holds for each flavor separately. 
The twist-3 distribution function $\tilde{g}_T(x)$ is the quark-gluon-quark correlation 
part of $g_T(x)=g_1(x)+g_2(x)$ split off by means of the equations of motion (for $m=0$): 
$\tilde{g}_T(x)=g_T(x)-g_{1T}^{(1)}(x)/x$ (cf.\ e.g.\ Refs.\ 
\cite{Tangerman:1994bb,Metz:2008ib,Accardi:2009au}). It can for instance be measured in the Drell-Yan 
process in double spin asymmetries $A_{LT}$ of longitudinally polarized hadrons colliding with transversely 
polarized hadrons \cite{Tangerman:1994bb,Boer:1997bw} or using SIDIS \cite{Accardi:2009au}.

Upon taking the lowest Mellin moment of Eq.\ (\ref{unintegratedrelation}), one arrives at:
\beq
\int_{-1}^1 {T^q(x,S_T)}\; dx = - c M^2 R_0 \left[ d_2^q +d_2^{\bar{q} }\right],
\eeq
where 
\beq
d_2^q = 3 \int_0^1 x^2
\left. g_2^q(x)\right|_{\text{{\small twist-3}}} dx = 2 \int_0^1 x^2\, 
\tilde{g}_T^q(x) \, dx.
\label{relationg2gTtilde}
\eeq
Summing over quark flavors, one obtains a relation in terms of the twist-3 part of the {\it structure} function $g_2$:
\beq
\sum_{q} e_q^2 \int_{-1}^1 {T^q(x,S_T)}\; dx = - 6 c M^2 R_0 \int_0^1 {x^2
\left. g_2(x)\right|_{\text{{\small twist-3}}}} \; dx,
\label{myESGM}
\eeq
which apart from the sum over flavors and the quark charge squared factor is a factor of 2 different from Eq.\ (\ref{ESGM}).
Given the uncertainty in the proportionality constant $c$ this factor is not of
importance, but we do note that the flavor dependence of the ESGM relation was not addressed properly in Ref.\ \cite{ESGM}.

The question here is whether $\tilde{g}_T(x)$ has a sign change 
as function of $x$, since the unintegrated ``ESGM" relation implies similar behavior for
$T(x,S_T)$. The bag model \cite{JaffeJi} suggests that $\left. g_2(x)\right|_{\text{{\small twist-3}}}$
is a sign changing function of $x$. However,  
${\left. g_2(x)\right|_{\text{{\small twist-3}}}}$ and ${\tilde{g}_T(x)} $ correspond to different operator matrix
elements, even though the second moments are directly related through Eq.\ (\ref{relationg2gTtilde}).
Instead, one can look at the first moment of $\tilde{g}_T(x)$, which in 
the $A^+=0$ gauge is given by: 
\beq
\int_{-1}^1 x\, 
\tilde{g}_T^q(x) \, dx = {\rm Re}\ \amp{P,S|  \psibar (0) {\gamma^\mu}  \gamma_5 
gA^{\nu} (0) \psi(0) | P,S} \frac{n_{-\mu} S_{T\nu}}{M} = 0,
\label{ELTme}
\eeq
where the vanishing of this particular matrix element is shown in Ref.\ \cite{Efremov:1996hd}
on the basis of Lorentz invariance, analogous to the derivation of the 
Burkhardt-Cottingham sum rule $\int_{-1}^1 g_2^q(x) \, dx =0$. Since the first moment of $\tilde{g}_T(x)$
deals with a local operator that appears in the Operator Product Expansion (OPE) for charged currents, one 
can more specifically conclude that also $\int_{0}^1 x\, \tilde{g}_T^q(x) \, dx$ corresponds to a local operator matrix element (obtained through Taylor expansion) that vanishes due to Lorentz invariance, 
in contrast to $\int_{0}^1 g_2^q(x) \, dx$ which does not appear in the local OPE \cite{Jaffe:1996zw,Blumlein:1996vs}. 
In other words, the vanishing integral in Eq.\ (\ref{ELTme}) is not due to a cancellation among the quark ($x>0$) and antiquark $(x<0$) 
contributions.

The vanishing of $\int_{0}^1 x\, \tilde{g}_T^q(x) \, dx$ implies that $\tilde{g}_T^q(x)$ has a node and 
through the unintegrated ``ESGM" relation also the QS function would have a node. 
In that case, $\int T(x,S_T) \; dx$ can be much smaller than the maximum value of $T(x,S_T)$, such that a sign 
change of $T(x,S_T)$ can accomodate both small $d_2$ and large SSA in a limited $x$ range
in a natural way. The unintegrated relation is useful in this respect, since asymmetries can be measured as 
function of $x$, therefore, the relation can be checked in experiment. The option that the QS function
has a node should then be kept in mind and also that the node can change position as a function of $Q^2$, 
which is relevant when comparing different experiments. 

What speaks against a node is that most model calculations of the Sivers function
\cite{Brodsky:2002cx,Yuan:2003wk,Bacchetta:2003rz,Cherednikov:2006zn,Gamberg:2007wm,Courtoy:2008dn,Pasquini:2010af}  
do not show a node (ignoring some small bag model artifacts), 
except for those of Refs.\ \cite{Lu:2004au,Courtoy:2008vi} which obtain down quark Sivers functions 
with a node, albeit very different ones, and for Ref.\ \cite{Bacchetta:2008af} which obtains an up 
quark Sivers function with a node. These model calculations of the Sivers functions all consider the
gauge link to lowest nontrivial order in the coupling constant, 
in other words, the first order expansion of the Wilson line. It
is unclear what the size of the higher order corrections
is and whether these could change the sign in a particular $x$ region. 

A common feature of the model results is that the up and down quark
Sivers functions are opposite in sign. This is also expected from the large
$N_c$ limit, which leads to $f_{1T}^{\perp u}(x,k_T^2)=-f_{1T}^{\perp d}(x,k_T^2) +{\cal O}(N_c^{-1})$ 
\cite{Pobylitsa:2003ty,Drago:2005gz}, and on the
basis of the signs of the quark anomalous magnetic moments through an
integral relation of the Sivers function and GPDs
\cite{Burkardt:2002ks,Burkardt:2007rv}. This suggests that if a node
occurs, it is present in both up and down quark Sivers functions.   

SIDIS experiments off proton
and deuteron targets in the valence region seem to indicate that up and down Sivers
functions are indeed opposite in sign and moreover similar in magnitude.
In some models the magnitude of the up quark
Sivers function was found to be much larger than that of the down
quark \cite{Bacchetta:2003rz,Lu:2004au}, but in more recent model
calculations they are found to be of comparable magnitude
\cite{Cherednikov:2006zn,Courtoy:2008vi,Bacchetta:2008af,Courtoy:2008dn,Pasquini:2010af}. 
If indeed of opposite sign, but similar in magnitude, the so-called  
Burkardt sum rule \cite{Burkardt:2004ur} 
\beq
\sum_{a=q,g} \int f_{1T}^{\perp (1) a}(x)\; dx = 0,
\eeq 
can be satisfied by a cancellation among contributions of the valence quarks, not requiring large contributions 
(and accompanying large cancellations) from non-valence contributions. This would be one possibility.
See Ref.\ \cite{Goeke:2006ef} for a check of the Burkardt sum rule in a model 
calculation that includes the gluon Sivers function.  A second possibility to satisfy the Burkardt sum rule is that the 
$x$ integral of the Sivers moment is zero for each flavor separately. The latter is
not in contradiction with the measured Sivers asymmetries in SIDIS, 
since those do not provide full information on the integral over $x$ of the Sivers moment. A future observation of a node 
in the Sivers function could point to this second possibility, as does a small gluon Sivers function. The latter can be 
seen by rewriting the Burkardt sum rule in the form:
\beq
\sum_{q} \int_{-1}^{1} T^q(x,S_T) \; dx  = - \int_{-1}^{1} T^g(x,S_T)\;  dx ,
\eeq
and by comparing it to the integrated ESGM relation of Eq.\ (\ref{myESGM}), where the l.h.s.\ of the latter includes an additional 
quark charge squared factor. If the gluon QS (or Sivers) function turns out to be small, then combined with the fact that 
$d_2$ is small, it implies that both $\sum_{q} \int_{-1}^{1} T^q(x,S_T) \; dx$ and $\sum_{q} e_q^2 \int_{-1}^{1} T^q(x,S_T) \; dx$ 
are small, disfavoring large cancellations among the flavors, but rather suggesting the $x$ integral of the QS function to be small 
for each flavor separately. In this way a small gluon QS/Sivers function and a node in the 
quark Sivers functions would go hand in hand. 

In conclusion, the $x$-dependent version of the ESGM relation presented here implies that the QS function changes sign 
as a function of $x$, but it is based on some assumptions that may not hold. It would be very 
interesting to test this relation in experiment, even outside the region of a possible node, since 
the proportionality relation should hold for all $x$ values. Confirmation of or deviations from the
relation could teach us about the qualitative features of the gluonic field strength inside the proton. 
Despite this uncertainty a node should be considered as a serious possibility, since twist-3 functions are 
not probability densities, therefore need not be positive definite. It would be just as natural for 
the QS function to have a node as it is for the twist-3 part of $g_2$. A sign change as function of $x$ also 
offers another way to satisfy the Burkardt sum rule, without requiring large cancellations among the different flavors. 
This sum rule together with the integrated ESGM relation obtained here suggest that a small gluon QS/Sivers 
function disfavors such large cancellations.   
A node is found in a few model calculations for the Sivers function, but not in most, and not 
for up and down quarks simultaneously, contrary to expectation from the large $N_c$ limit. 
Extending these model calculations of the Sivers function to include its 
full Wilson line, or equivalently performing model calculations of the Qiu-Sterman function, would therefore 
be very interesting.
  
\begin{acknowledgments}
I thank Mauro Anselmino, Umberto D'Alesio, Xiaodong Jiang, Francesco Murgia, Alexei Prokudin, Oleg Teryaev and Werner Vogelsang for useful discussions. 
\end {acknowledgments} 


\end{document}